\documentclass[aip,apl,reprint,a4paper,superscriptaddress,floatfix,amsmath,amssymb,amsfonts,noshowpacs,longbibliography]{revtex4-2}

\usepackage{newtxtext,newtxmath}
\usepackage[utf8]{inputenx}
\input{ix-utf8enc.dfu}
\usepackage{graphicx}
\usepackage[dvipsnames]{xcolor}
\usepackage{soul} 
\usepackage{xspace}
\usepackage{siunitx}

\usepackage[
,textwidth=17.5cm
,textheight=23.7cm
,verbose
,pdftex
]{geometry}
\usepackage{xspace}

\usepackage[pdftex]{hyperref}
\hypersetup{
	colorlinks,
	citecolor=blue,
	linkcolor=blue,
	urlcolor=blue}

\newcommand{\iga}{$\mathrm{In}_{x}\mathrm{Ga}_{1-x}\mathrm{N}$\xspace}
\newcommand{\igt}{In\textsubscript{0.12}Ga\textsubscript{0.88}N\xspace}

\begin{document}
\graphicspath{{./figures/}}

\title{Growth of compositionally uniform  In$_x$Ga$_{1-x}$N layers with low relaxation degree on GaN by molecular beam epitaxy }

\author{Jingxuan Kang}
\author{Mikel Gómez Ruiz}
\author{Duc Van Dinh}
\author{Aidan F Campbell}
\author{Philipp John}
\author{Thomas Auzelle}
\author{Achim Trampert}
\author{Jonas Lähnemann}
\author{Oliver Brandt}
\author{Lutz Geelhaar}
\email[Electronic email: ]{Geelhaar@pdi-berlin.de}
\affiliation{Paul-Drude-Institut für Festkörperelektronik, Leibniz-Institut im Forschungsverbund Berlin e.V., Hausvogteiplatz 5--7, 10117 Berlin, Germany.}

\begin{abstract}

500-nm-thick \iga layers with $x=$ 0.05--0.14 are grown using plasma-assisted molecular beam epitaxy, and their properties are assessed by a comprehensive analysis involving x-ray diffraction, secondary ion mass spectrometry, and cathodoluminescence as well as photoluminescence spectroscopy. We demonstrate low degrees of strain relaxation (10\% for $x=0.12$), low threading dislocation densities ($\mathrm{1\times10^{9}\,cm^{-2}}$ for $x=0.12$),  uniform composition both in the growth and lateral direction, and a narrow emission band. The unique sum of excellent materials properties make these layers an attractive basis for the top-down fabrication of ternary nanowires. 

\end{abstract}
                 
\maketitle

The ternary alloy \iga is arguably the most important optoelectronic material since it covers the wide spectral range from the near-infrared to the entire visible region and has enabled ubiquitous solid-state lighting.\cite{yamInGaNOverviewGrowth2008a} In particular, \iga quantum wells emitting blue light exhibit internal quantum efficiencies approaching unity thanks to mature growth procedures.\cite{nakamuraCandelaClassHigh1994} At the same time, the level of material quality is still substantially lower for quantum wells with higher In content and layers with a thickness exceeding a few hundred nanometers. The central challenge in this context is the lattice mismatch to GaN templates and the lack of better-suited substrates. Specifically, during the growth of thick layers with low to moderate In content, strain relaxes gradually which in turn leads to increasing In incorporation, a phenomenon called \lq{compositional pulling}\rq.\cite{schulzInfluenceStrainIndium2020a,pereiraCompositionalPullingEffects2001a} Moreover, strain relaxation is associated with the formation of threading dislocations (TDs) which act as non-radiative recombination centers. \cite{speckRoleThreadingDislocations1999a,lahnemannCarrierDiffusionMathrm2022} \par

The growth of \iga layers with thickness on the order of hundreds of nanometers is mainly motivated by two purposes: First, such layers could serve as pseudo-substrates for the growth of \iga quantum wells with high In content emitting, e.\,g., in the red spectral region.\cite{evenEnhancedIncorporationFull2017a,emaGrowthLatticerelaxedInGaN2021a} Second, such layers could be used as active region in optoelectronic converters such as photodetectors and solar cells.\cite{bhuiyanInGaNSolarCells2012a,kongRecentProgressInGaNbased2022} While both purposes require thick layers, the required properties differ significantly. The use as pseudo-substrate necessitates a high degree of strain relaxation to provide a template with low mismatch to the desired quantum wells with high In content. In contrast, for the active region of light-absorbing devices, \cite{fabienIIINitrideDoubleHeterojunctionSolar2016,fengTheoreticalSimulationsEffects2010a, zhangEffectDislocationsEfficiency2013a} a low degree of strain relaxation is crucial to avoid TDs and other defects reducing quantum efficiency.\cite{lahnemannCarrierDiffusionMathrm2022} Furthermore, in the second case a spatially highly homogeneous In content is typically desirable, whereas compositional gradients in the growth direction do not impede the functionality of pseudo-substrates and may actually be introduced on purpose.

A third, new motivation for thick \iga layers is to employ them as the starting point for the top-down fabrication of \iga nanowires. The nanowire geometry offers conceptual advantages for optoelectronic applications based on light absorption such as photocatalytic water splitting.\cite{chatterjeeIIInitrideNanowiresSolar2017a} In particular, light can be absorbed more efficiently than in thin layers, and charge carriers can be collected in the direction orthogonal to the impinging light, thus reducing the distance from the active region. However, in the bottom-up growth of \iga nanowires, the resulting morphology is strongly coupled to the chosen composition and doping.\cite{kamimuraSiDopingEffects2014a, zhangInGaNNanowiresHigh2016a} In contrast, in a top-down approach the former is defined during etching, and the latter during initial planar epitaxy.

The layer requirements for subsequent top-down nanowire fabrication resemble those of layers serving as light-absorbing active region in that compositional uniformity is desirable and structural defects should be avoided. In particular, top-down nanowires could inherit TDs from the original layer, which would significantly reduce the nanowire quantum efficiency. Under the assumption that the TDs are randomly distributed in the layer, the number $n$ of TDs in top-down nanowires is statistically determined by the density $\rho$ of TDs in the layer  as well as the nanowire diameter $d$: $n = d^2\pi\rho/4$. Thus, it is expected that a planar \iga layer with TD density of approximately $\mathrm{1\times10^{9}\,cm^{-2}}$ will yield a nanowire ensemble comprising $\geq$98\% TD-free nanowires if $d\leq50$\,nm.\cite{wangTopFabricationCharacterization2014} 

This study targets the growth of thick \iga layers suitable for the subsequent top-down fabrication into nanowires. We employ plasma-assisted molecular beam epitaxy (PA-MBE) to grow \iga layers with a thickness of 500\,nm and In contents of up to 0.14. Systematic examination of their structural and optical properties demonstrates a low relaxation degree and correspondingly low density of TDs, uniform composition, and a narrow photoluminescence (PL) emission band. \par

\iga layers with a thickness of 500 nm were grown using PA-MBE on commercial GaN(0001) templates at a substrate temperature of 620\,$\mathrm{^\circ C}$. Two In cells were employed, one of which served to compensate In desorption and maintain intermediate In-rich growth conditions, the other one to incorporate In for the desired stoichiometry. Before growth, In cell~A was set to a flux of $\mathrm{3.2 \times 10^{14}\,cm^{-2}s^{-1}}$, sufficient for the accumulation of In droplets on the surface. The time necessary for droplet formation was monitored by reflection high-energy electron diffraction (RHEED),\cite{brandtGaAdsorptionDesorption2004a} as explained in more detail in the supplementary material, and used as an indicator for the actual substrate temperature.\cite{gallinatGrowthDiagramPlasmaassisted2007a} Next, the flux of In cell~A was adjusted to ensure intermediate In-rich growth conditions.\cite{gallinatGrowthDiagramPlasmaassisted2007a} The cell shutter was opened and closed again after 60\,s, and the In cell~A temperature was fine-tuned utilizing RHEED intensity transients to ensure the formation of an In adlayer.\cite{brandtGaAdsorptionDesorption2004a} After these adjustments and full In desorption, all four shutters were opened. For the growth of \igt, the fluxes of In cell~B, Ga, and N were $\mathrm{5.8\times 10^{13}\,cm^{-2}s^{-1}}$, $\mathrm{5.5\times 10^{14}\,cm^{-2}s^{-1}}$, and $\mathrm{6.1\times 10^{14}\,cm^{-2}s^{-1}}$, respectively. In terms of the \iga layer stoichiometry, in addition to the In cell~B flux, the decomposition of \iga during the epitaxy process, which depends on the growth temperature as well as In composition, also affects the final In content of the epitaxial layers.\cite{nathMolecularBeamEpitaxy2010} For \iga layers with different In concentration, the metal flux of In cell~B and Ga was adjusted accordingly. \par

After growth, the morphology of the \iga layers was evaluated using RHEED and atomic force microscopy (AFM). Subsequently, a triple-axis high resolution x-ray diffractometer (Philips PANalytical X’Pert PRO MRD) equipped with a two-bounce hybrid Ge(220) monochromator and a $\mathrm{CuK}_{\alpha1 }$ source ($\lambda =1.540598\,\mathrm{\AA}$) was employed to reveal the structural properties of the \iga layers. The In content and strain relaxation degree were determined based on $2\theta-\omega$ scans of the 0002, 0004, $10\bar{1}2$, $10\bar{1}3$, $11\bar{2}2$, and $20\bar{2}1$ reflections as well as reciprocal space maps (RSM) around the $10\bar{1}5$ reflection.\cite{schusterDeterminationChemicalComposition1999b, moramXrayDiffractionIIInitrides2009} Moreover, to evaluate the structural quality of the \iga layers, rocking curves were acquired around the 0002 and $10\bar{1}2$ reflections. The TD density was extracted separately from maps of the panchromatic cathodoluminescence (CL) intensity obtained at 10\,K and transmission electron microscopy (TEM) measurements. Hyperspectral CL maps and secondary ion mass spectrometry (SIMS) were utilized to evaluate the compositional uniformity. Quantitative information about the In content was obtained from the SIMS measurements by comparison with a reference \iga layer whose composition was calibrated using Rutherford backscattering spetrometry. PL spectra were measured at room temperature using a Renishaw InVia setup to investigate the emission properties. The samples were excited by a 325\,nm He-Cd laser with a spot diameter of about 1\,{\textmu}m and an excitation density of $\approx$ 1 kW $\mathrm{cm^{-2}}$.\par

\begin{figure}[t]
	\centering
	\includegraphics[width=7.5cm]{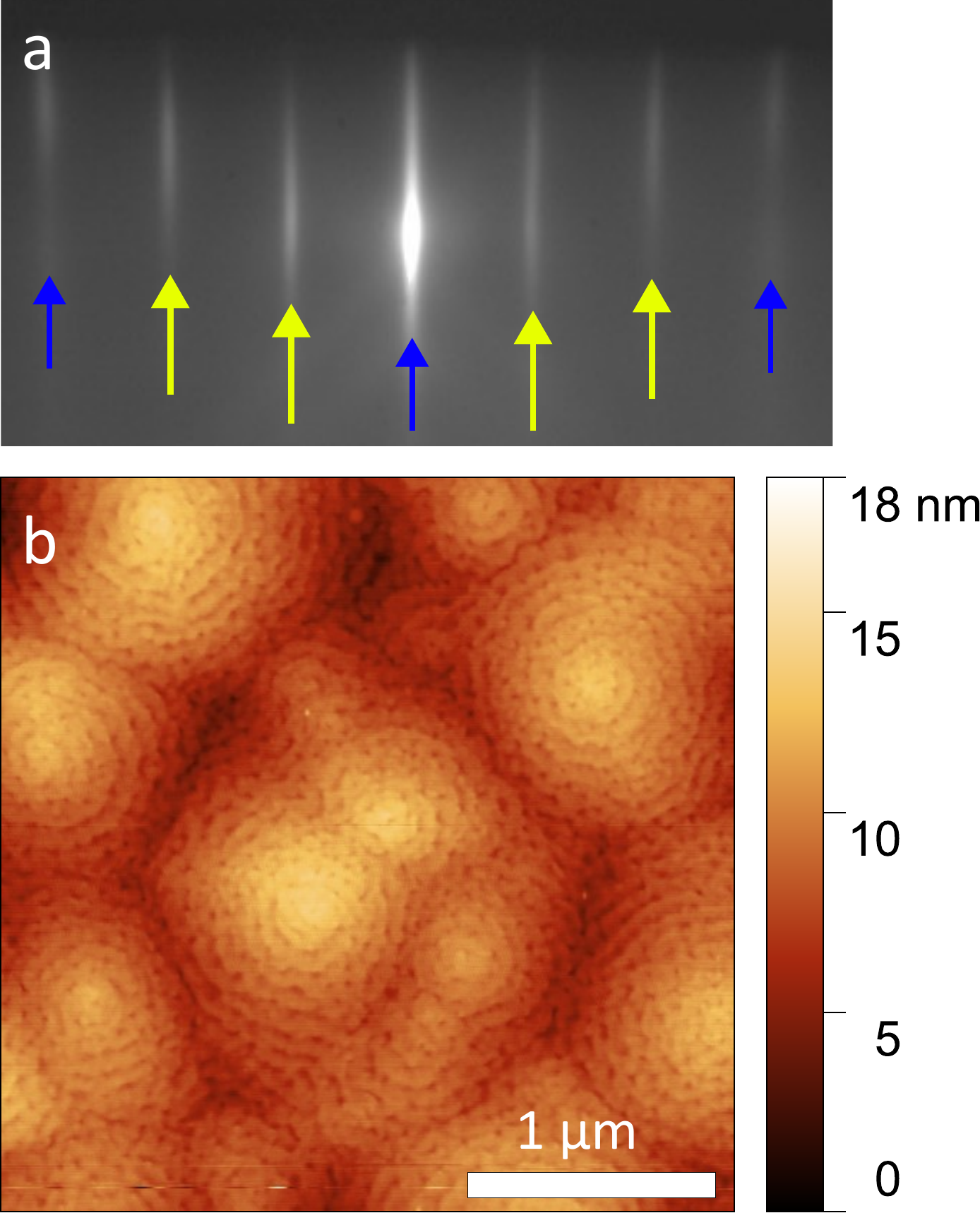}
	\caption{(a) RHEED pattern of the \igt surface after growth along the [1$\bar1$00] azimuth showing  the $\sqrt{3}\times\sqrt{3}\,\mathrm{R}30^{\circ}$ reconstruction induced by the In adlayer with integer and fractional reflections marked by blue and yellow arrows, respectively. (b)~AFM topograph of the \igt layer.}
	\label{fig:RHEED and AFM}
\end{figure}

The RHEED pattern of the \igt layer after growth presented in Fig.~\ref{fig:RHEED and AFM}(a) reveals the $\sqrt{3}\times\sqrt{3}\,\mathrm{R}30^{\circ}$ surface reconstruction known to reflect an In-terminated surface \cite{chen2000,chezeGaN0001302017} and exhibits sharp streaks that indicate a smooth surface. Consistently, the corresponding AFM topograph in Fig.~\ref{fig:RHEED and AFM}(b) shows the typical hexagonal pyramids characteristic for \iga grown by PA-MBE.\cite{hestrofferRelaxedCplaneInGaN2015a, hestrofferPlasmaassistedMolecularBeam2016a} The monolayer spiral growth around screw threading dislocations is well resolved, and the root-mean-square roughness extracted over a scanning area of \mbox{3$\times$3\,\textmu$\mathrm{m^{2}}$} is 1.9\,nm. The smooth surface, absence of $\vee$-pits, and the observation of the $\sqrt{3}\times\sqrt{3}\,\mathrm{R}30^{\circ}$ reconstruction confirm that the \iga layer was grown under In-stable conditions with an In adlayer with $>$ 1\,monolayer coverage.\cite{valdueza-felipHighIncontentInGaN2014a} \par

\begin{figure}[t]
	\includegraphics[width=\columnwidth]{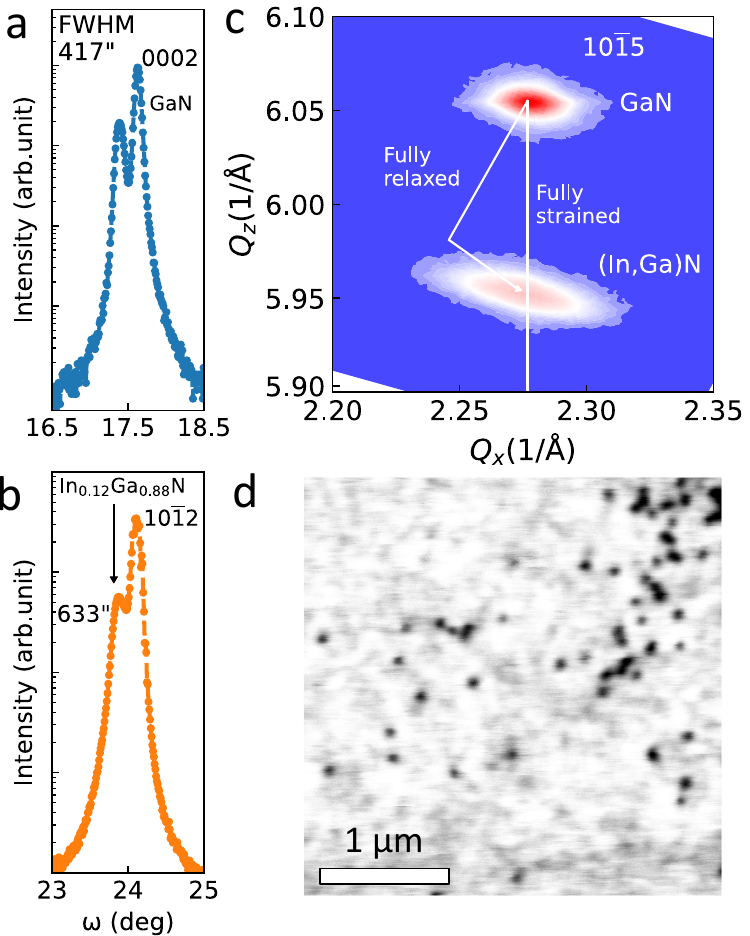}
	\caption{XRD $\omega$ scans of the \igt sample around the (a) 0002 and (b) $10\bar{1}2$ reflections performed with an open detector. (c) RSM around the GaN\,10\=15 reflection revealing an In content of $0.12$. (d) Panchromatic CL intensity map acquired at 10\,K. The dark spots correspond to TDs.}
	\label{fig:In-content}
\end{figure}

Figure~\ref{fig:In-content}(a) and \ref{fig:In-content}(b) display for the \igt layer x-ray diffraction (XRD) $\omega$ scans of the 0002 and $10\bar{1}2$ reflection.The profiles exhibit full widths at half maximum (FWHM) of $417^{\prime \prime}$ and $633^{\prime \prime}$, respectively. These FWHM values are larger than the GaN template, which are $286^{\prime \prime}$ and $449^{\prime \prime}$ respectively, yet are comparable to the FWHM reported for PA-MBE grown \iga ($x = 0.12$) on a GaN template, \cite{baziotiDefectsStrainRelaxation2015} as well as for strain-relaxed \iga pseudo-substrate with lower In content $x < 0.1$ obtained from strained \iga layer transfer \cite{evenEnhancedIncorporationFull2017a} with thicknesses less than the 500\,nm of our \igt layer. The RSM in Fig.~\ref{fig:In-content}(c) reveals for this layer $x = 0.12$ and a relaxation degree of only $10\%$. This In content and strain relaxation degree are consistent with the results calculated from the average lattice constant which is determined from $2\theta-\omega$ scans of the 0002, 0004, $10\bar{1}2$, $10\bar{1}3$, $11\bar{2}2$, and $20\bar{2}1$ reflections.\cite{schusterDeterminationChemicalComposition1999b} The 10\% strain relaxation degree of our \igt layer is on the lower end compared to reported values for epitaxial \iga layers with similar thickness and In content grown on GaN templates.\cite{monetaPeculiaritiesPlasticRelaxation2018a, baziotiDefectsStrainRelaxation2015, lvSurfaceEvolutionThick2021a}  \par

Direct insight into the TD density is provided by panchromatic CL intensity maps as depicted in Fig.~\ref{fig:In-content}(d). In addition to minor fluctuations across the surface, there are dark spots indicating positions of strongly reduced intensity that are the outcrops of TDs at the surface. \cite{lahnemannCarrierDiffusionMathrm2022} Averaging over different locations on the \igt layer, we obtained a TD density of $\mathrm{\approx 1\times10^{9}\,\mathrm{cm^{-2}}}$. For comparison, the TD density of the GaN (0001) template, determined using the same method before growth, is $\mathrm{\approx 2\times10^{8}\,\mathrm{cm^{-2}}}$. The TD density of the \igt layer was confirmed by transmission electron microscopy measurements (not shown). We note that the TD density is as low as the reported values for similar layers in  literature \cite{wangInvestigationStrainRelaxation2010, monetaPeculiaritiesPlasticRelaxation2018a} and is consistent with the low strain relaxation degree determined by XRD. \par 

\begin{figure}[t]
	\centering
	\includegraphics[width=\columnwidth]{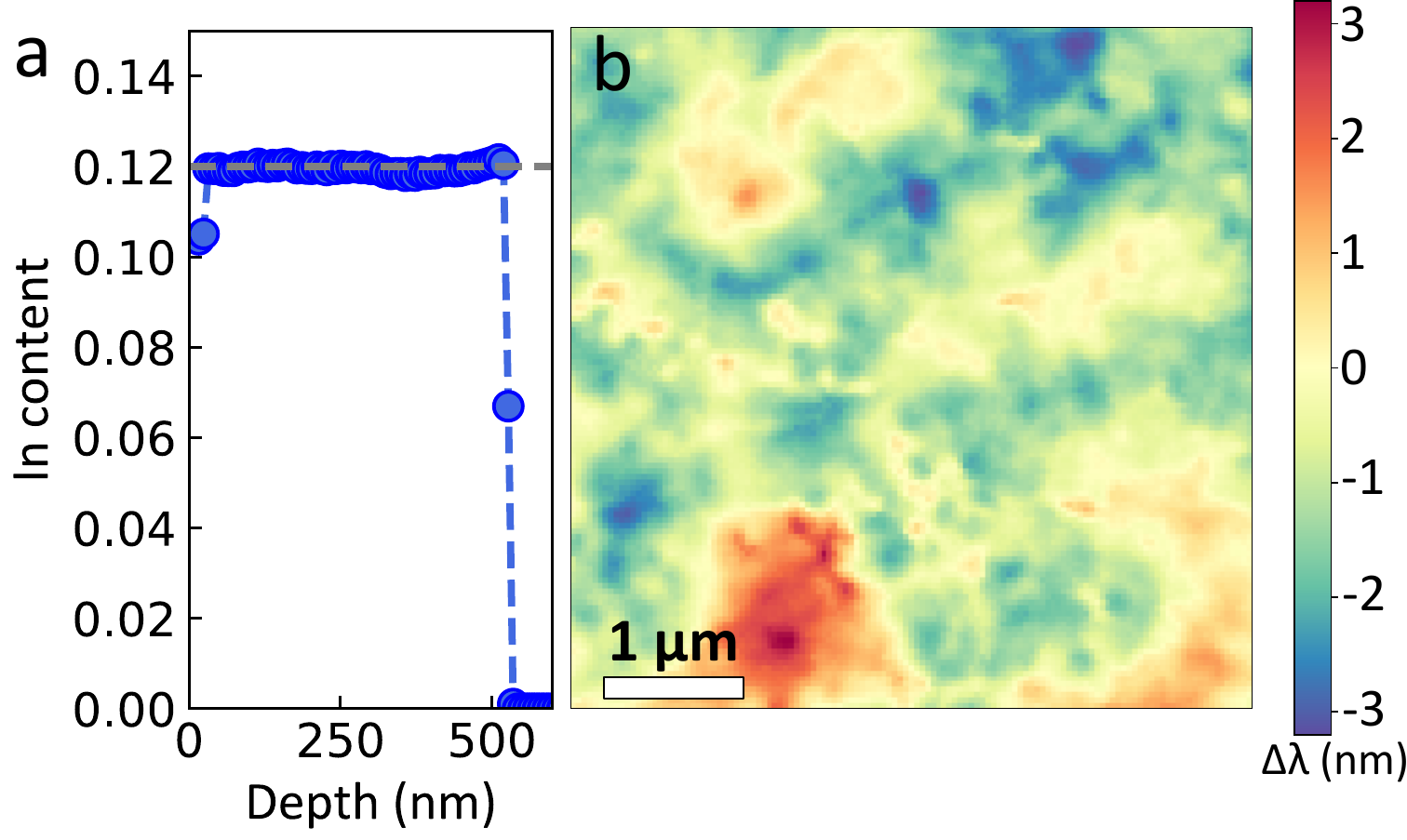}
	\caption{(a) In content depth profile obtained by SIMS. (b) Room temperature map of the variation of the CL peak emission wavelength.} 
	\label{fig:SIMS+CL}
\end{figure}

The spatial homogeneity of the In content is another factor critical for \iga to be used as active region in devices. A depth profile of the In content in our \igt layer was obtained by SIMS, and the result is presented in Fig.~\ref{fig:SIMS+CL}(a). The In content remains constant along the growth direction throughout the entire layer. In contrast, during the growth of thick \iga layers the In content often increases due to compositional pulling.\cite{pereiraCompositionalPullingEffects2001a,baziotiDefectsStrainRelaxation2015} The high uniformity in the growth direction found here is a direct consequence of the low strain relaxation degree of our \igt layer. To assess the compositional homogeneity in the lateral direction, maps of the CL peak emission wavelength were acquired, as depicted in Fig.~\ref{fig:SIMS+CL}(b). Within this area of 5 $\times$ 5 \textmu m$^2$, the variation of the peak emission wavelength amounts to $\pm 3$\,nm, indicating that there are only negligible fluctuations in In content on the local scale ($< 0.8\%$). The combination of SIMS and hyperspectral CL maps demonstrates that our \igt layer exhibits a uniform In content both in the growth and lateral direction. \par
 
\begin{figure}[t]
	\centering
	\includegraphics[width=6cm]{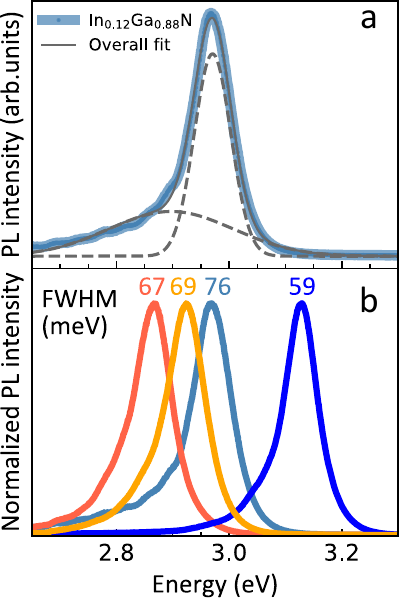}
	\caption{(a) Room temperature PL spectrum (broad turquoise line) of the \igt layer fitted (grey solid line) with two Gaussians (dashed grey lines). (b) Room temperature PL spectra of thick \iga layers with In content ranging from 0.05 to 0.14.}
	\label{fig:PL summary}
\end{figure}

Room temperature PL spectroscopy is another way to assess the compositional uniformity of the \iga samples. The spectrum of the \igt layer is depicted in Fig.~\ref{fig:PL summary}(a). The spectrum is comparably narrow but its shape indicates the presence of two distinct emission bands. Fitting with two Gaussians results in peak energies of 2.97 and 2.89\,eV. The band at 2.97\,eV corresponds to the recombination of (presumabely localized) excitons, while the other one is attributed to the first LO replica. Note that the spectral separation of the zero phonon line and the first LO replica does not equal the LO energy due to their different density of states, which introduces a shift proprtional to kT. \cite{Bebb1972} The FWHM of the excitonic emission resulting from the Gaussian fit is 76\,meV, suggesting a low degree of alloy disorder within the sample.\cite{shanOpticalPropertiesInxGa11998,weisbuchDisorderEffectsNitride2021}  \par

So far, we have focused on the \iga layer with 0.12 In content. In addition, a series of samples with measured In content spanning from 0.05 to 0.14 was grown under identical conditions, the only variation being the In/Ga flux ratio. The corresponding PL spectra are shown in Fig.~\ref{fig:PL summary}(b) and reveal a monotonic red shift from 3.13 to 2.87\,eV with increasing In content. In parallel, the FWHM values are all in range 60 to 76\,meV. Data for the structural analysis of the sample whose PL emission is centered at 2.87\,eV is presented in the supporting information. In short, we extracted an In content of 0.14, a strain relaxation degree of 20\%, a TD density of $1.9 \times10^{9}\,\mathrm{cm^{-2}}$, and an almost uniform In content in the growth direction.  \par

Compared to the \igt layer, the higher In content of 0.14 leads to more plastic strain relaxation and accordingly a higher TD density. At the same time, the FWHMs of the PL spectra do not differ much between the different samples. In general, spectral broadening is attributed to \iga alloy disorder.\cite{shanOpticalPropertiesInxGa11998, weisbuchDisorderEffectsNitride2021} Thus, our sample series feature very uniform composition. We emphasize that for all of our samples the PL FWHMs are comparable to the best values reported in literature for layers with similar In contents and thicknesses.\cite{nakamuraHighQualityInGaNFilms1992a, arteevInvestigationStatisticalBroadening2018a, liliental-weberStructuralPerfectionInGaN2009a} \par

In summary and conclusion, we have demonstrated the growth of 500-nm-thick \iga layers with In contents up to 0.14 that combine various desirable properties. In particular, the relaxation degree and the TD density are low, the compositional uniformity is high, and the PL emission band is narrow. These layers present an excellent starting point for the top-down fabrication of \iga nanowire ensembles with highly uniform properties and high quantum efficiency. It is expected that $\geq 96\%$ of top-down nanowires fabricated from the \iga layer with $x =0.14$ are free of TDs if the mean nanowire diameter is 50\,nm. \par

The authors thank Hans-Peter Schönherr for technical support and Andrea Ardenghi for critically reading the manuscript. This work has been supported by Deutsche Forschungsgemeinschaft under grants Ge2224/6-1 and Au610/1-1.\par

\bibliography{ms}

\newpage

\end{document}


\graphicspath{{./figures/}}

\title{Growth of compositionally uniform  \iga layers with low relaxation degree on GaN by molecular beam epitaxy }

\author{Jingxuan Kang}
\author{Mikel Gómez Ruiz}
\author{Duc Van Dinh}
\author{Aidan F Campbell}
\author{Philipp John}
\author{Thomas Auzelle}
\author{Achim Trampert}
\author{Jonas Lähnemann}
\author{Oliver Brandt}
\author{Lutz Geelhaar}
\affiliation{Paul-Drude-Institut für Festkörperelektronik, Leibniz-Institut im Forschungsverbund Berlin e.V., Hausvogteiplatz 5--7, 10117 Berlin, Germany.}

\maketitle

\subsection*{\fontsize{13}{16}\selectfont Supplementary Material}

\subsection{Substrate temperature calibration by reflection high-energy electron diffraction}

\begin{figure*}
\centering
\includegraphics[width=14cm]{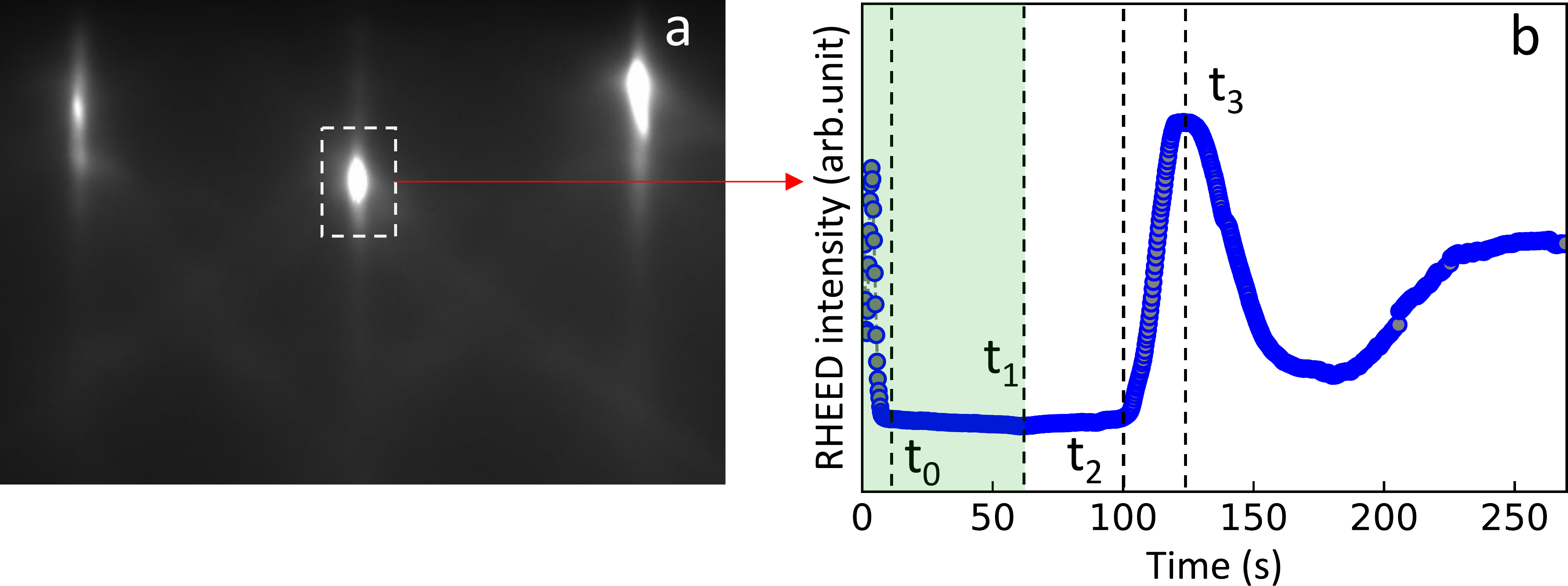}
\caption{(a) RHEED pattern of the GaN (0001) substrate surface along the [1$\bar1$00] azimuth. The dashed rectangle marks the area around the specular spot whose intensity evolution during the temperature calibration process is shown in (b). The In shutter is open during the time indicated by the green shading. The times $t_0$, $t_1$, $t_2$, and $t_3$ mark the completion of the In adlayer, the closing of the In shutter, the full desorption of the In droplets, and an oscillation associated with the desorption of the In adlayer.}
\label{RHEED}
\end{figure*}

Prior to growth, the substrate temperature is calibrated by determining the desorption rate of In droplets. To this end, the intensity of the specular spot in the reflection high-energy electron diffraction (RHEED) pattern is recorded, as illustrated in Fig.~\ref{RHEED}(a) and reported previously.\cite{brandtGaAdsorptionDesorption2004a} A fixed In flux $\Phi_0$ is provided which is sufficiently high to enable the accumulation of In droplets on the GaN(0001) surface. The corresponding evolution of the RHEED intensity is depicted in Fig.~\ref{RHEED}(b). Upon opening of the In cell shutter at the time 0, the RHEED intensity inceases first and then steadily decreases during the formation of an In adlayer. Following the completion of the In adlayer at the time $t_0$, excess In accumulates into droplets on top of the adlayer while the RHEED intensity remains constant. At the time $t_1$ (60\,s in this study), the In cell shutter is closed, and the droplets are getting consumed by replenishing the adlayer, which is continuously desorbing. The RHEED intensity remains constant until the droplets have vanished. Subsequently, at time $t_2$ the adlayer desorbs and the RHEED intensity increases. During the adlayer desorption process, intensity oscillations may occur (as indicated by $t_3$) if a bilayer has formed. The mass balance for In droplet formation and desorption on the adlayer allows to determine the desorption rate $\Phi_{des}$ from equation \ref{T_cali} below. In turn, the In desorption rate is governed by the substrate temperature, and for our calibration we use the data reported by Gallinat et al.\cite{gallinatGrowthDiagramPlasmaassisted2007a}\par

\begin{equation}
\Phi_{des}=\Phi_0 \times \frac{t_1-t_0}{t_2-t_0}
\label{T_cali}
\end{equation}

\subsection{Structural analysis of the \igf layer}

\begin{figure*}
\centering
\includegraphics[width=15cm]{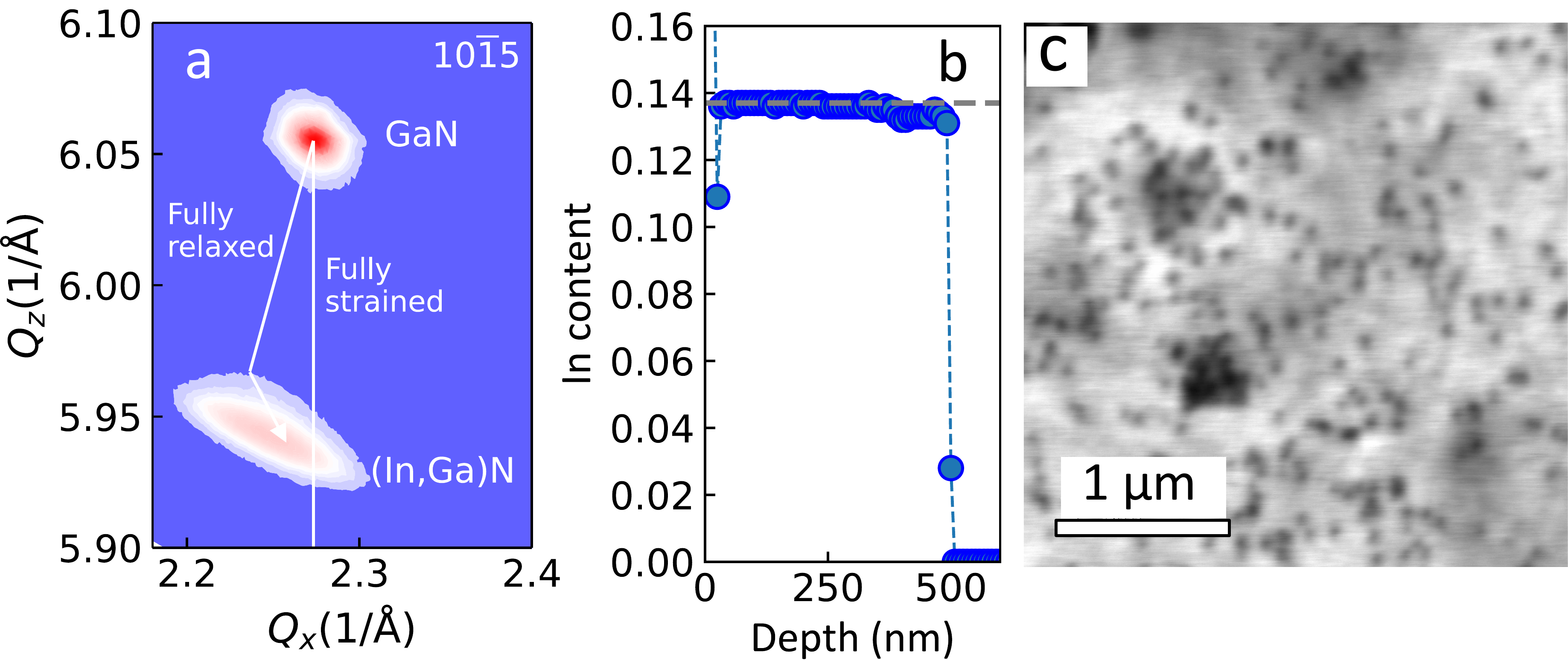}
\caption{ Structural analysis of the \igf layer. (a) RSM around the GaN\,10\=15 reflection. (b) In content depth profile obtained by SIMS. (c) Panchromatic CL intensity map acquired at 10\,K.}
\label{M1817}
\end{figure*}

The structural analysis of the \igf layer  corresponding to the photoluminescence spectrum centered at 2.87\,eV in the main text was carried out in the same way as presented in detail there for the \igt layer. The results are shown in Fig.~\ref{M1817}. The reciprocal space map (RSM) in Fig.~\ref{M1817}(a) reveals an In content of 0.14 and a strain relaxation degree of 20\%. These values are confirmed from $2\theta-\omega$ scans at  the 0002, 0004, $10\bar{1}2$, $10\bar{1}3$, $11\bar{2}2$, and $20\bar{2}1$ reflections. The In content profile in Fig.~\ref{M1817}(b), measured by secondary ion mass spectrometry (SIMS), reveals an  In content of 0.14 in accordance with the RSM result. A slight increase is observed after the growth of the first 100\,nm of \iga. This increase is attributed to strain relaxation. The dark spots on the panchromatic cathodoluminescence (CL) spectroscopy intensity map in Fig.~\ref{M1817}(c) correspond to a threading dislocation density of approximately $1.9 \times10^{9}\,\mathrm{cm^{-2}}$.\par

\bibliography{supplement}